\documentclass[twocolumn,showpacs,superscriptaddress,preprintnumbers,floatfix]{revtex4}
\usepackage{graphicx}

\def\be{\begin{equation}}
\def\ee{\end{equation}}
\def\bea{\begin{eqnarray}}
\def\eea{\end{eqnarray}}
\begin{document}

%\preprint{gr-qc/yymmddd}

\title{A Remark on "A CMB/Dark Energy Cosmic Duality"}

\author{Fabio Finelli}
\author{Alessandro Gruppuso}
\affiliation{INAF/IASF, Istituto di Astrofisica Spaziale e
Fisica Cosmica, Sezione di Bologna \\
via Gobetti, 101 -- I-40129 Bologna -- Italy}

\date{\today}

\begin{abstract}
The recent calculation on the suppression of the power at low
multipoles in the CMB spectrum due to an IR cut-off presented in
hep-th/0406019 does not take into account the Integrated Sachs-Wolfe
(ISW) term, which is crucial in models aiming to the explanation of the
present acceleration of the Universe. We show that the ISW contribution
to low multipoles is tipically much greater than the SW term,
for an IR cut-off comparable to the present Hubble radius.
\end{abstract}


\maketitle

%\section*{Introduction}
%{\bf 1.} 

\vskip 0.5cm

%\noindent

Recently an attempt to lower the quadrupole of the CMB temperature 
anisotropies pattern by imposing an IR cut-off has been tried in 
\cite{ES}.
The IR cut-off is imposed for a flat universe in the relation between the
multipoles of the
CMB anisotropy pattern and scalar metric fluctuations in Fourier space.
Such relation for an adiabatic scale invariant spectrum is
\cite{KS}:
\begin{equation}
\langle \left(\frac{\Delta T}{T} \right)^2 \rangle_{\ell}=
\frac{A^2}{100 \, \pi \, \ell(\ell+1)} K^2_{\ell} \ ,
\label{ISWlambda}
\end{equation}
where $A$ is the amplitude of gravitational fluctuations and
the coefficient $K^2_l$ is given by \cite{KS}
\begin{eqnarray}
K_\ell^2 &=& 200 \, \ell(\ell+1) \int_{k_c}^{\infty}
\frac{dk}{k} \left[{1 \over 10} j_\ell(k(\eta_0 - \eta_r)) + \right.
\nonumber \\
%\nonumber\\
&& \left. + \int_{\eta_r}^{\eta_0} d\eta \, \frac{d f}{d \eta}
j_\ell(k(\eta_0-\eta)) \right]^2 \nonumber \\
& \equiv &
%200 \, \ell(\ell+1)
\int_{s_c}^{\infty} d s
%\frac{dk}{k}
\left[ I_{\ell \, \rm SW} (s) + I_{\ell \, \rm ISW} (s)
\right]^2
\label{Kell}
\end{eqnarray}
where $s=k/H_0$, $\eta_r \,, \eta_0$ are the conformal times at
recombination and at
present, respectively.
$k_c$ is an IR cut-off and the function $f$, defined as
\be
f(\eta) = 1-{a^{\prime} \over a^3} \int_0^{\eta} d \tau \, a^2(\tau)
\, ,
\label{sol}
\ee
describes the time-dependence of metric fluctuations in a
$\Lambda$CDM scenario.

The ISW term is zero in a CDM scenario (where $K_\ell = 1$), while
%It was shown almost twenty years ago that the ISW effect was
is actually of the same order of magnitude of the SW term at low $\ell$ in
presence of $\Lambda$ \cite{KS}.
All the results presented here are obtained with
$\Omega_\Lambda^0=0.75$, $\Omega_b^0=0.05$, $\Omega_{\rm CDM}^0=0.2$,
$h=0.7$ (the same as in \cite{ES}).

\begin{figure}

\includegraphics[scale=0.7]{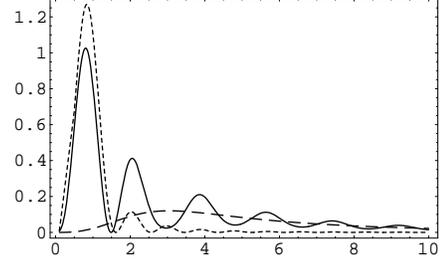}

\caption{$I_{2 \, \rm SW}^2$ (short-dashed), $I_{2 \, \rm ISW}^2$
(long-dashed) and the total $(I_{2 \, \rm SW} + I_{2 \, \rm ISW})^2$
(solid) with respect to $s=k/H_0$. The IR cut-off used in
\cite{ES} corresponds to $s_* \simeq 2.6$.}
\label{fig1}

\end{figure}

The integrands of the SW
and ISW contributions look very different,
as shown in Fig. 1 for $\ell=2$. An IR cut-off can easily kill
the SW term, but not the ISW term since its shape is broader in
Fourier space.
%Therefore the ISW contribution to the quadrupole (and lower
%multipoles) cannot be much decreased by an IR cut-off.
Table 1 shows how
the SW term is reduced by two orders of magnitude,
while the ISW term is almost unchanged.

\begin{table}[t]

\begin{center}
\begin{tabular}{|c|c|c|c|}
\hline
$s_c=k_c/H_0$ & $  K_2^2  $ & SW &  ISW \\
\hline
$ \pi 10/12 $ & $0.500 $ & $0.035$ & $0.594$ \\
\hline
 $1/10$ & $1.544$ & $0.999$ & $0.600$ \\
\hline

\end{tabular}
\end{center}
\caption{Numerical values of $K_2^2$ (and the relative SW and ISW
contributions) for the IR cut-off chosen in
\cite{ES} and for a much smaller cut-off (not zero for numerical
reason).}

\end{table}

Fig. 2 shows the CMB temperature power spectrum computed by CMBFAST
\cite{LOS}.
Note that the introduction of an IR cut-off of the same magnitude as in
\cite{ES} reduces the low tail of the CMB power spectrum approximately to
one third of the $\Lambda$CDM value, in agreement with Table 1
and in contrast with Fig. 1 of \cite{ES}. Our result is also in agreement
with \cite{contaldi}.

We finally note that, for the cut-off chosen in \cite{ES}, Fig. 1 shows
that the ISW term is larger than the SW term, {\em independently} from the discretization
procedure used in \cite{ES}, which substitutes the integral in Eq. (\ref{Kell}) with a
sum as in Eq. (6) of \cite{ES}.

\begin{figure}

\includegraphics[scale=0.3]{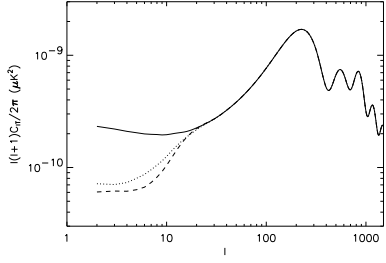}

\caption{CMB temperature power spectrum computed with CMBFAST: the solid
line is a $\Lambda$CDM model without IR cutoff, the dashed line has an IR
cut-off, the dotted line has the same IR cut-off, but a different window
function.}
\label{fig2}

\end{figure}

%Similar considerations hold for dynamical DE models 
%(but with additional DE perturbations).

\end{document}